\documentclass[10pt,twocolumn,preprintnumbers,superscriptaddress,nofootinbib,aps,prl]{revtex4}

\usepackage{graphicx}
\usepackage{amsmath}
\usepackage{srcltx}
\usepackage[utf8]{inputenc}
\usepackage[colorlinks]{hyperref}
\usepackage{times}
\usepackage{subfigure}
\newcommand{\vev}[1]{\langle {#1} \rangle}
\newcommand{\lsim}{\lesssim}

\newcommand{\ord}[1]{\mathcal{O}{(#1)}}
\newcommand{\beq}{\begin{equation}}
\newcommand{\eeq}{\end{equation}}

\newcommand{\scond}{\vev{\bar s \, s}}

\newcommand{\appropto}{\mathrel{\vcenter{
  \offinterlineskip\halign{\hfil$##$\cr
    \propto\cr\noalign{\kern2pt}\sim\cr\noalign{\kern-2pt}}}}}

\begin{document}

\pagestyle{plain}

\title{\boldmath 
Good things to do with extra Higgs doublets
}

\author{Hooman Davoudiasl\footnote{email: hooman@bnl.gov}
}

\affiliation{High Energy Theory Group, Physics Department, Brookhaven National Laboratory,
Upton, New York 11973, USA}

\author{Ian M. Lewis\footnote{email: ian.lewis@ku.edu}
}

\affiliation{Department of Physics and Astronomy, University of Kansas, 
Lawrence, Kansas, 66045 USA}

\author{Matthew Sullivan\footnote{email: msullivan1@bnl.gov
}
}

\affiliation{High Energy Theory Group, Physics Department, Brookhaven National Laboratory,
Upton, New York 11973, USA}


\begin{abstract}
{\bf Executive Summary}: In this contribution to the Snowmass 2021 process, we outline  models with two or three Higgs doublets that address open  questions of particle physics and cosmology.  In particular, we show that with two additional Higgs doublets one can provide a mechanism for the generation of lepton asymmetry and hence baryon asymmetry, through CP violating  Higgs decays, near weak scale temperatures.  In another model with only one extra Higgs doublet, we illustrate that Yukawa couplings to quarks and neutrinos can lead to a viable mechanism for the generation of Dirac neutrino masses, sourced by the QCD chiral condensate of strange quarks.  We adapt  Spontaneous Flavor Violation -- a framework  for coupling light fermions to new Higgs doublets while avoiding tree level flavor-changing neutral currents -- in constructing these models.  
In both cases, flavor data provide interesting constraints on the parameter space.  Either scenario includes $\ord{1}$ couplings of light quarks to the Higgs doublets which allow a future 100 TeV $pp$ collider to have reach for the new scalars up to $\ord{10~{\rm TeV}}$ masses, through resonant single production.  In the neutrino mass model, collider data can shed light on the mass hierarchy of neutrinos.  This article is based on work presented in Refs.~\cite{Davoudiasl:2021syn,Davoudiasl:2021nfv}.

\end{abstract}
\maketitle

\section{Introduction}
 Interactions with light fields often allow for new physics 
to be probed at current and future collider experiments and hence lead to  accessible phenomenology.   
In this article, we will summarize our proposals \cite{Davoudiasl:2021syn,Davoudiasl:2021nfv} 
for how such setups can be motivated by models that address key open problems of particle physics and cosmology.  
In particular, we will focus on explaining the baryon asymmetry of the Universe (BAU) and  non-zero masses for the SM neutrinos with additional heavy Higgs doublets.  Our models involve $\ord{1}$ coupling of light quarks to the new scalars, providing significant reach for the heavy states, through resonant single production at hadron colliders.  However, 
any proposal for  new physics close to the weak scale needs to comply with stringent bounds on flavor-changing neutral currents (FCNCs) \cite{Bona:2007vi}.  In the case of models with more than one Higgs doublet, the Spontaneous 
Flavor Violation (SFV) framework \cite{Egana-Ugrinovic:2018znw,Egana-Ugrinovic:2019dqu} provides a prescription for how to couple the extra 
scalars to light Standard Model (SM) quarks and leptons, 
in a way that avoids tree level FCNCs.  Let us then begin with a  brief outline of the SFV framework employed in our models.


\section{Spontaneous Flavor Violation}

The SFV \cite{Egana-Ugrinovic:2018znw,Egana-Ugrinovic:2019dqu} framework is a paradigm that mitigates unwanted flavor violating effects and allows for more freedom in Yukawa structure than the typical Minimal Flavor Violation. In particular, the light quarks can have large couplings to heavy Higgs doublets in the SFV framework \cite{Egana-Ugrinovic:2019dqu}. There are up-type and down-type SFV schemes; for our purposes, we will use the up-type option. We will slightly generalize this into two models which we will call Model A and Model B, for simplicity. Model A will have three Higgs doublets, while Model B will have two Higgs doublets.

For $N$ Higgs doublets, the general Yukawa sector including right-handed neutrinos would look like 
\begin{eqnarray}
&\displaystyle\sum_{a=1}^N&-\lambda^a_u \bar Q \, \epsilon \, H_a^* \,u - \lambda^a_d \bar Q \, H_a \, d - 
\lambda^a_\nu \bar L \, \epsilon\, H_a^* \,\nu_R - \lambda^a_\ell \bar L \, H_a \,\ell\,\nonumber\\
&&+{\rm H.C.}
\label{eq:Yukawa}
\end{eqnarray}  
We will always work in the Higgs basis, where only the first doublet, $H_1$, gets a vacuum expectation value. In the up-type SFV 2HDM with no right-handed neutrinos, these Yukawa coupling matrices take the form
\begin{equation}
\begin{array}{ccc}
\lambda^1_u = V_{CKM}^\dagger Y_u,& \quad \lambda^1_d = Y_d,& \quad  \lambda^1_\ell = Y_\ell  \\
\lambda^2_u = \xi V_{CKM}^\dagger Y_u,& \quad \lambda^2_d = K_d,& \quad \lambda^2_\ell = \xi_\ell Y_\ell ,
\label{eq:uptypeSFV}
\end{array}
\end{equation}
where $Y_u$, $Y_d$, and $Y_\ell$ are the diagonal SM Yukawa couplings in the mass basis for up-type quarks, down-type quarks, and charged leptons, respectively.  Here, $V_{CKM}$ is the CKM matrix, $K_d=diag(\kappa_d,\kappa_s,\kappa_b)$ is flavor diagonal with real entries $\kappa_{d,s,b}$; $\xi$ and $\xi_\ell$ are real constants. This will be the basis of our flavor structure for Model A and Model B. We will minimally generalize this to three Higgs doublets in Model A by having the same couplings $\lambda^3_x=\lambda^2_x$ for $x=u,d,\ell$.

Since we add right-handed neutrinos, we also need a suitable generalization of the SFV paradigm to the neutrino sector. A natural generalization in the same spirit as SFV would be
\begin{eqnarray}
\lambda^1_\nu &=& V_{PMNS} Y_\nu \nonumber \\
\lambda^2_\nu &=& V_{PMNS} K_\nu ,\label{eq:lamnu}
\end{eqnarray}
where $K_\nu=diag(\kappa_{\nu,1}, \kappa_{\nu,2}, \kappa_{\nu,3})$. In Model A, $Y_\nu$ will be the diagonal Yukawa couplings in the mass basis which generate the neutrino masses; in Model B, we will have $Y_\nu=0$ and generate masses in another way. Model B will use $\kappa_{\nu,i}$ real. For Model A, we will similarly duplicate the same sort of Yukawa structure to a third Higgs doublet, with a notable change: we will impose $|\lambda^3_\nu|=|\lambda^2_\nu|$, but will allow arbitrary CP phases, different between the $H_2$ and $H_3$ couplings.

\section{Flavor Constraints}

\begin{figure*}[tb]
\begin{center}
\subfigure[]{ \includegraphics[width=\columnwidth]{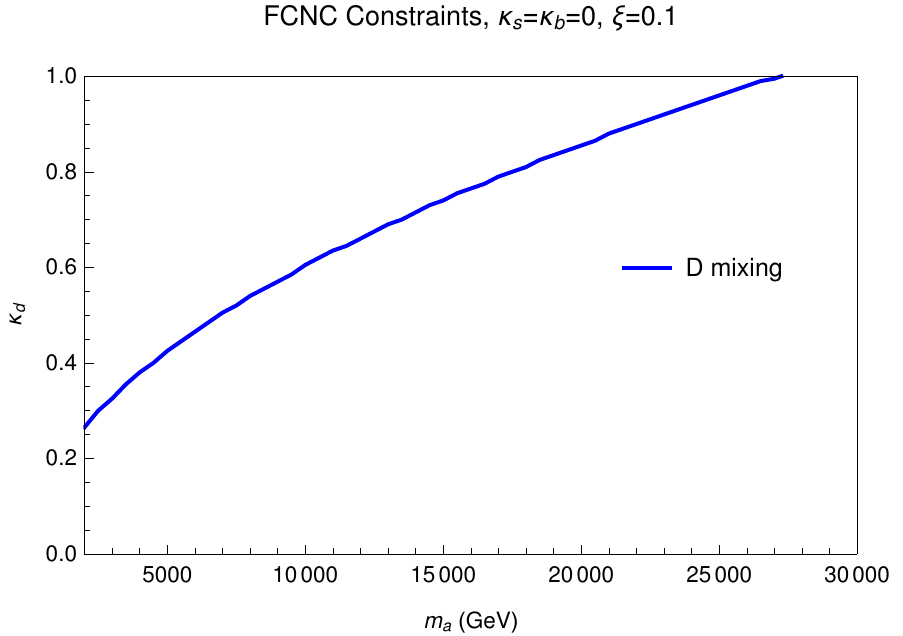}}
\subfigure[]{ \includegraphics[width=\columnwidth]{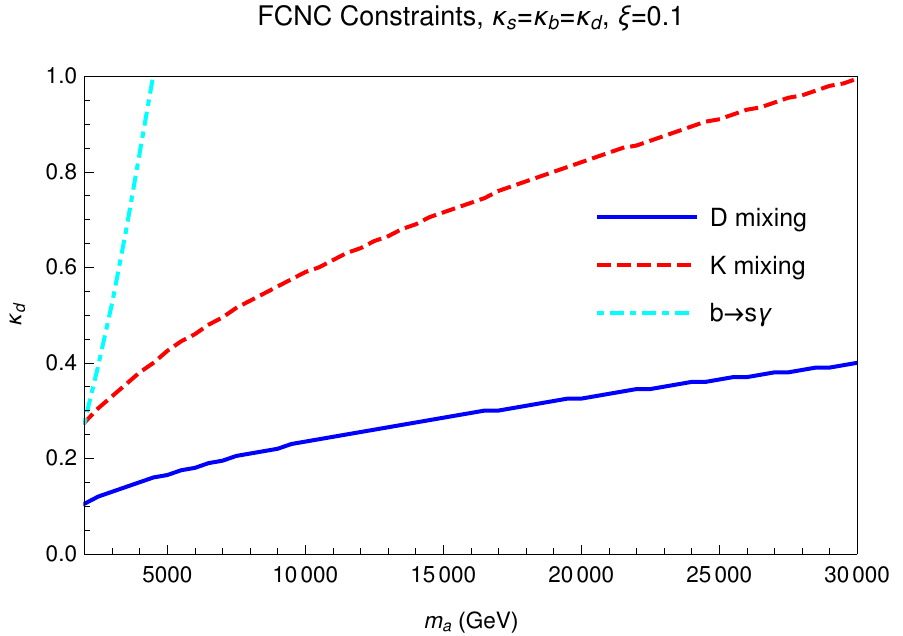}}
\caption{\label{fig:BaryoFlavor} Constraints on Model A from flavor changing neutral currents for $\xi=0.1$ and the different flavor structures (a) $\kappa_s=\kappa_b=0$, (b) $\kappa_s=\kappa_b=\kappa_d$. Measurements mentioned in the text that are not shown on the plot are not constraining in the parameter range shown.}
\end{center}
\end{figure*}

Though the SFV scheme suppresses FCNC processes, there are still experimental bounds which are constraining for certain regions of parameter space, particularly with lighter Higgs masses. There are flavor-changing quark decays as well as meson mixing bounds to consider. We considered experimental bounds coming from the flavor-changing decays $b \to d\gamma$ \cite{Crivellin:2011ba} and $b\to s\gamma (\ell^+\ell^-)$ \cite{Capdevila:2017bsm}, as well as neutral meson mixing for $K-\bar{K}$\cite{Bona:2007vi}, $B_d-\bar{B}_d$ \cite{Bona:2016bvr}, $B_s-\bar{B}_s$ \cite{Bona:2016bvr}, and $D-\bar{D}$ \cite{Aaij:2019jot}. We used the formulas presented in Ref.~\cite{Egana-Ugrinovic:2019dqu} to calculate the contributions to these processes. We show the relevant limits that these experimental bounds place on the couplings in Model A, with two degenerate heavy Higgs doublets, in Fig.~\ref{fig:BaryoFlavor}.

In Model B, for reasons that we will discuss later, we will only have strange quark couplings and neutrino couplings for the heavy doublet. This makes many of the flavor bounds less important. Particularly, only $D-\bar{D}$ mixing constraint was relevant. We use the recently updated measurements from Ref.~\cite{LHCb:2021ykz} to constrain the parameter space of this model, though the constraint on $\kappa_s$ is only slightly different from the previous measurement in Ref.~\cite{Aaij:2019jot}.



\section{Higgs Troika Baryogenesis}

\begin{figure}[tb]
\begin{center}
\includegraphics[width=0.5\textwidth,clip]{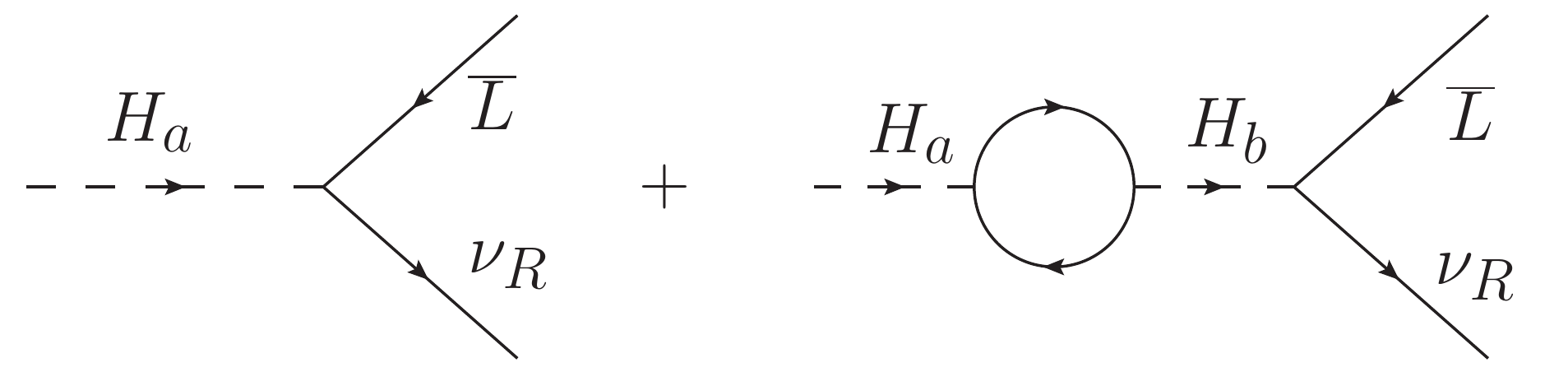}
\caption{\label{fig:BaryoDiag} Feynman diagrams illustrating the Higgs doublet decays that generate the BAU}
\end{center}
\end{figure}

In this section, we review the baryogenesis mechanism proposed in Ref.~\cite{Davoudiasl:2021syn}, which used a three Higgs doublet model -- the {\it Higgs Troika} -- and the interactions given by Eqs.~(\ref{eq:Yukawa}),(\ref{eq:uptypeSFV}), assuming  Dirac neutrinos.  This is Model A described earlier.  The BAU is generated via the Higgs doublet decays into a lepton doublet and right-handed neutrino, as shown in Fig.~\ref{fig:BaryoDiag}.  In order to generate the Dirac neutrino masses, the SM-like Higgs has neutrino Yukawas of the order $\mathcal{O}(10^{-12})$.  If the SM-like Higgs participated in the decays in Fig.~\ref{fig:BaryoDiag}, this coupling would be far too weak and the BAU would not be generated.  Hence, we need two additional Higgs doublets that can couple more strongly to the neutrinos.

\begin{figure*}[tb]
\begin{center}
\subfigure[]{  \includegraphics[width=0.45\textwidth]{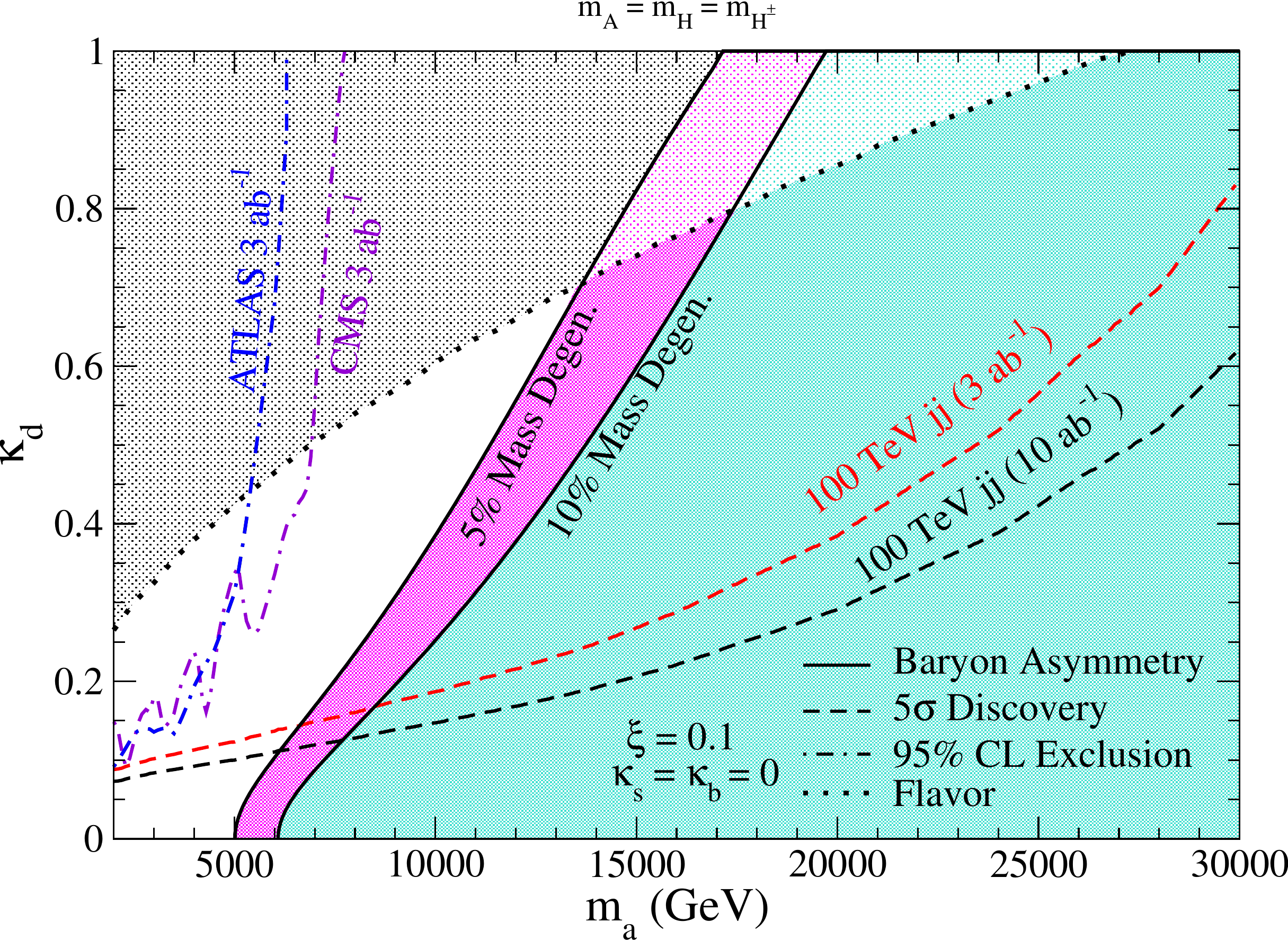}}
\subfigure[]{  \includegraphics[width=0.45\textwidth]{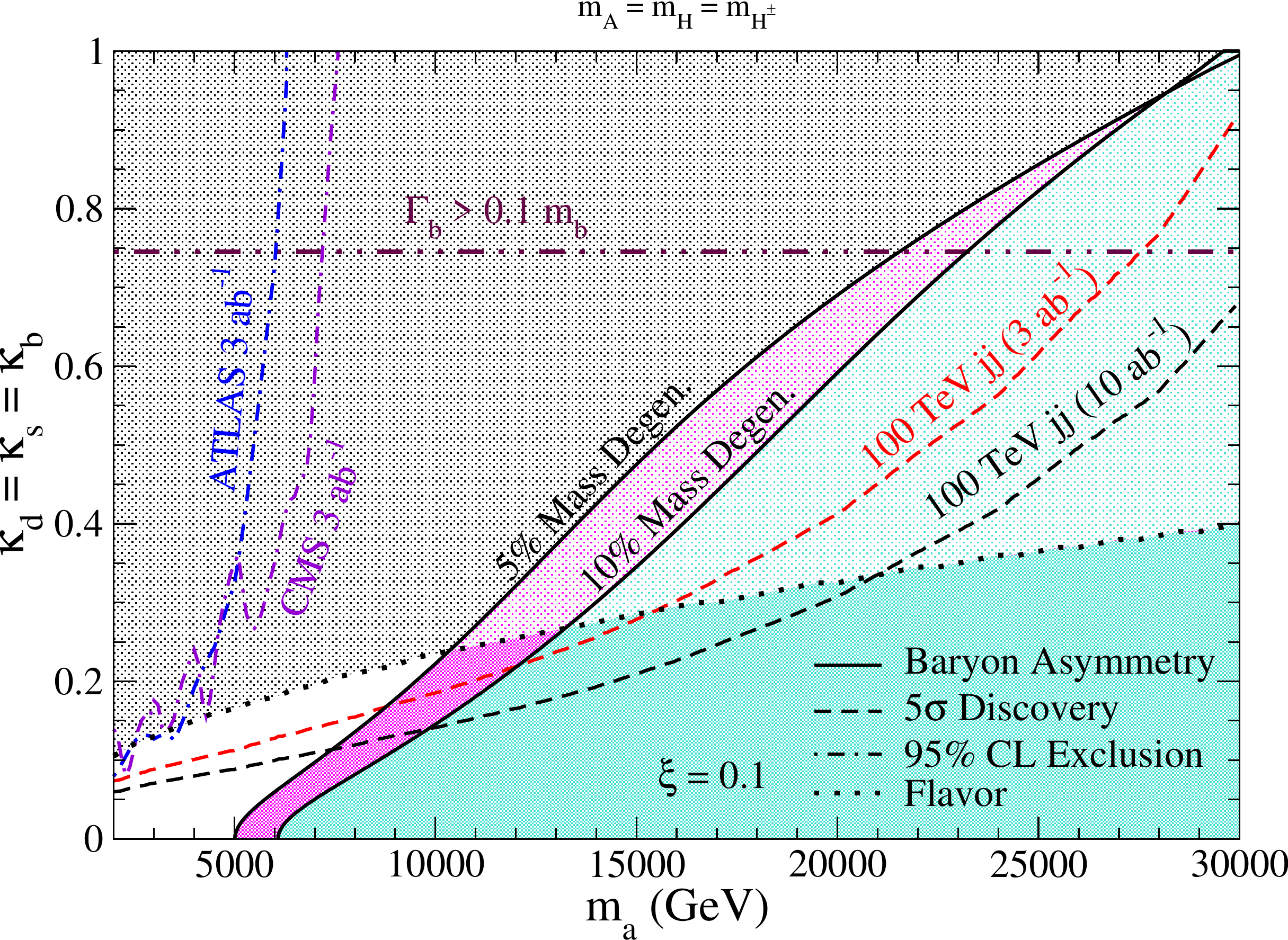}}
\caption{\label{fig:BaryoCollider} Regions of parameter space for which the BAU can be generated with (shaded magenta) a 5\% mass degeneracy $m_a/m_b=0.95$ between the two Higgs doublets and (shaded turqoise) a 10\% mass degeneracy $m_a/m_b=0.9$.  The discovery reach at a 100 TeV $pp$ collider in the dijet channel is above the dashed lines for (red) 3 ab$^{-1}$ and (black) 10 ab$^{-1}$.  Regions above the dot-dashed lines can be exlcuded by (blue) ATLAS~\cite{Aad:2019hjw} and (violet) CMS~\cite{Sirunyan:2019vgj} with 3 ab$^{-1}$.  The gray shaded regions above the dotted lines are ruled out by flavor.  In (b) the region above the dot-dot-dashed lines have a Higgs width larger than 10\% of its mass.  In (a) we set $\xi=0.1,\,\kappa_s=\kappa_b=0$ and in (b) we set $\xi=0.1,\kappa_d=\kappa_s=\kappa_b$.  In both figures we consider all components of the Higgs doublet to be degenerate.}
\end{center}
\end{figure*}

The BAU is generated by creating an asymmetry in the lepton doublets that is then processed into a baryon asymmetry via electroweak sphalerons. Hence, we need a reheat temperature of $T_{rh}\sim 100$~GeV for the sphalerons to be in thermal equilibrium.  For Higgs fields with masses $\mathcal{O}(1~{\rm TeV})$, the reheat temperature cannot be much higher or else the scalars would be in thermal equilibrium with the plasma and any asymmetry would be washed out.\footnote{The original population of the heavy Higgs fields could be sourced via the decays of a modulus $\Phi$ as detailed in Ref.~\cite{Davoudiasl:2019lcg,Davoudiasl:2021syn}.}  Even though the Higgs doublets themselves may not be in thermal equilibrium, they can still mediate processes $Ff\leftrightarrow L\nu_R$, where $F$ ($f$) are SM SU(2) doublet (singlet) fermions.  Requiring that the rates of these processes are less than the rate of expansion of the Universe at $T_{rh}\sim100~{\rm GeV}$ constrains the Yukawa couplings of the two heavy Higgs boson~\cite{Davoudiasl:2021syn}:
\beq
\lambda_\nu^a \,\lambda_f^a  \lsim 2.1 \times 10^{-4} \left(\frac{m_a}{10~\text{TeV}}\right)^2\,,
\label{Ha-washout}
\eeq 

In order to generate the observed BAU \cite{Tanabashi:2018oca} 
\beq
\frac{n_B}{s} \approx 9\times 10^{-11}\,,
\label{obs-BAU}
\eeq 
we need an asymmetry parameter~\cite{Davoudiasl:2019lcg,Davoudiasl:2021syn}
\begin{eqnarray}
\varepsilon_a=\frac{\Gamma(H_a\rightarrow \bar{L}\nu_r)-\Gamma(H_a^\star\rightarrow L\bar{\nu_R})}{2\Gamma(H_a)}\gtrsim  10^{-7},\label{eq:AsymBnd}
\end{eqnarray}
where $\Gamma(H_a)$ is the total width of $H_a$ and we assumed the mass of the modulus that sources the Higgs population is 60 TeV or larger.  

Including Higgs width effects~\cite{Pilaftsis:1997dr}, the asymmetry parameter can be calculated to be
\begin{eqnarray}
\varepsilon_a = \frac{1}{8\pi}\frac{(m_b^2-m_a^2)m_a^2}{(m_b^2-m_a^2)^2 + m_b^2\Gamma_b^2}\frac{\sum_{f=q} N_{c,f}\text{Im}\left({\rm Tr}^{ba}_\nu {\rm Tr}^{ba*}_{f}\right)}{\sum_{f=q}N_{c,f}{\rm Tr}^{aa}_f},
\label{eq:eps}
\end{eqnarray}
where ${\rm Tr}^{ba}_f={\rm Tr}[\lambda^{b\dagger}_{f}\lambda^a_f]$ and $a,b$ indices correspond to the two heavy Higgs doublets.
When the flavor structure of Eqs.~(\ref{eq:uptypeSFV},\ref{eq:lamnu}) is combined with the washout bound of Eq.~(\ref{Ha-washout}), there is an upper bound on the asymmetry parameter
\begin{eqnarray}
\varepsilon_a &\lesssim& 1.8\times10^{-9}\left(\frac{m_a}{10 {\rm TeV}}\right)^4\frac{(m_b^2/m_a^2-1)}{(m_b^2/m_a^2-1)^2 + m_b^2\Gamma_b^2/m_a^4}\nonumber\\
&&\times \frac{1}{\kappa_d^2+\kappa_s^2+\kappa_d^2+\xi^2}\,.\label{eq:epsnum}
\end{eqnarray}
Simultaneously requiring $\varepsilon_a\gtrsim 10^{-7}$ from Eq.~(\ref{eq:AsymBnd}), limits the parameter region that can realistically generate the BAU.  The parameter range is shown as the magenta and cyan shaded regions below the solid black lines in Fig.~\ref{fig:BaryoCollider}.  The magenta region corresponds to a $5\%$ mass degeneracy between the two heavy Higgs doublets, and the cyan region corresponds to a $10\%$ mass degeneracy.  As can be clearly seen, the BAU can be generated for heavy Higgs  masses $m_a\gtrsim 5-10$~TeV.

As can be seen in Figs.~\ref{fig:BaryoFlavor} and \ref{fig:BaryoCollider}, the flavor and BAU constraints allow non-negligible couplings between the heavy Higgses and the light quarks.  With these couplings, the main production mode of the Higgses at hadron colliders is single production via initial state quark anti-quark annihilation: $q\bar{q}\rightarrow A,H,H^\pm$, where $A,H,H^\pm$ are the neutral pseudoscalar, neutral scalar, and charged scalar components of a Higgs doublet.  If the quark couplings are non-neglible, which is also needed to singly produce multi-TeV Higgses, the washout condition of Eq.~(\ref{Ha-washout}) requires that the neutrino coupling $\lambda_\nu^a$ to be relatively small compared to the quark Yukawas.  Hence, the main decay modes of the heavy Higgses will be to dijets.

In Fig.~\ref{fig:BaryoCollider}, the regions above the dashed lines can be discovered at a 100 TeV $pp$ collider with (red) 3 ab$^{-1}$ and (black) 10 ab$^{-1}$ respectively.  These discovery reaches were found by adapting the di-jet discovery reach of a $Z'$ model with quark only couplings from Ref.~\cite{Golling:2016gvc};  details of the adaptation method can be found in Ref.~\cite{Davoudiasl:2021syn}.  We see that a 100 TeV $pp$ collider is sensitive to a large portion of parameter space with viable baryogenesis.  The regions above the dot-dash lines can be ruled out at the LHC at 95\% CL with 3 ab$^{-1}$ of data.

For the discovery reach we assumed one Higgs doublet whose components are degenerate in mass such that new physics gives contributions to oblique parameters that vanish~\cite{Peskin:1990zt,Peskin:1991sw,Barbieri:2006dq,Haber:2010bw,Ahriche:2015mea} and contributions to the electric dipole moments that are negligible~\cite{Pilaftsis:1997dr,Davoudiasl:2021syn}.  The production of the different components were added incoherently.  Due to CP violation, there will be some interference between the neutral scalar and pseudoscalar production~\cite{Pilaftsis:1997dr}.  However, we expect that interference to be small.  Another source of interference is between the two Higgs doublets.  If the mass difference between the two doublets is less than the Higgs decay widths, we could expect coherent enhancements up to a factor of two~\cite{Davoudiasl:2021syn}.  Indeed, the BAU is enhanced by a degeneracy between the Higgs doublets.

\section{Strange Neutrino Masses}

\begin{figure*}[tb]
\begin{center}
\subfigure[]{\includegraphics[width=0.45\textwidth,clip]{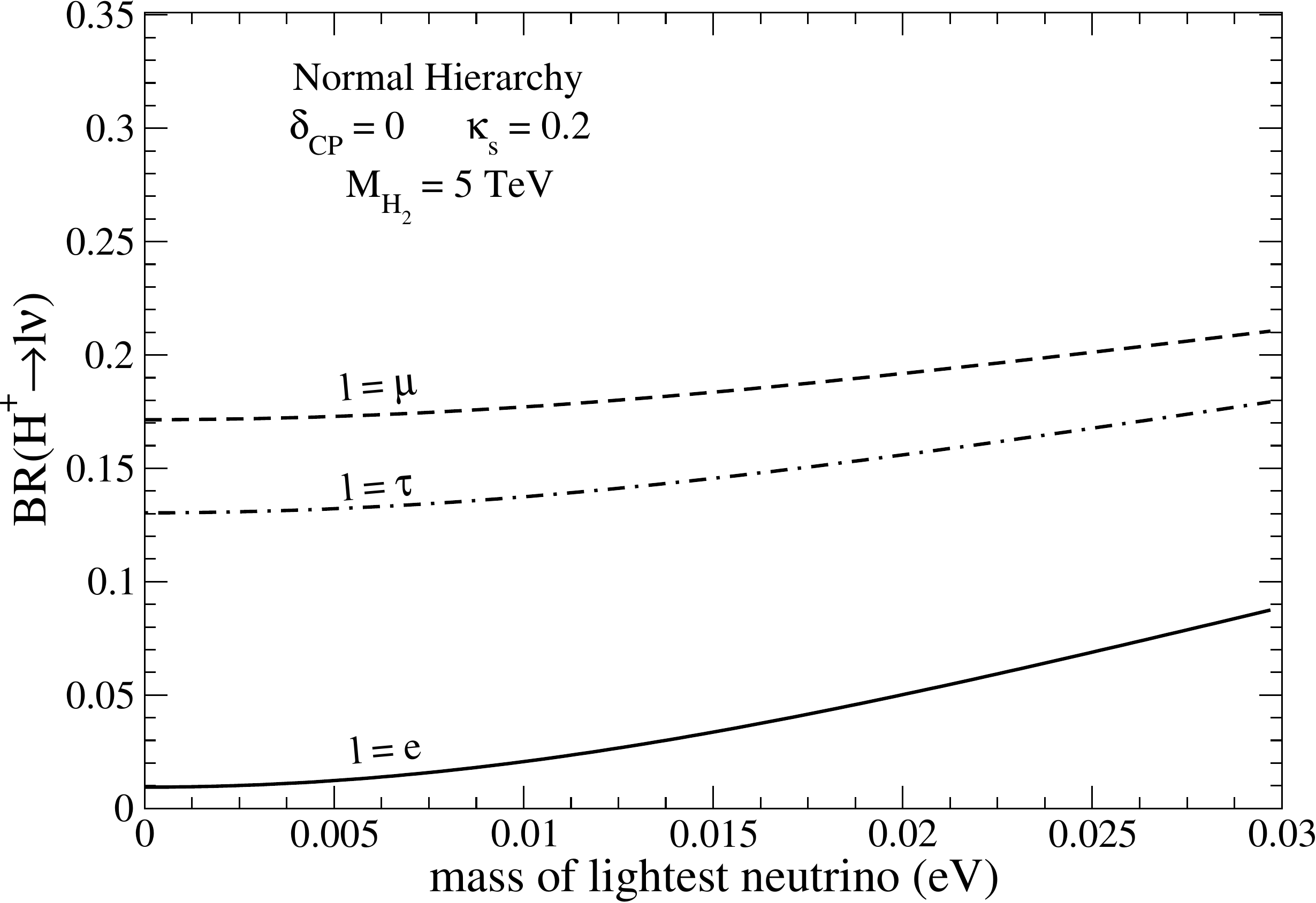}\label{fig:BR_norm}}
\subfigure[]{\includegraphics[width=0.45\textwidth,clip]{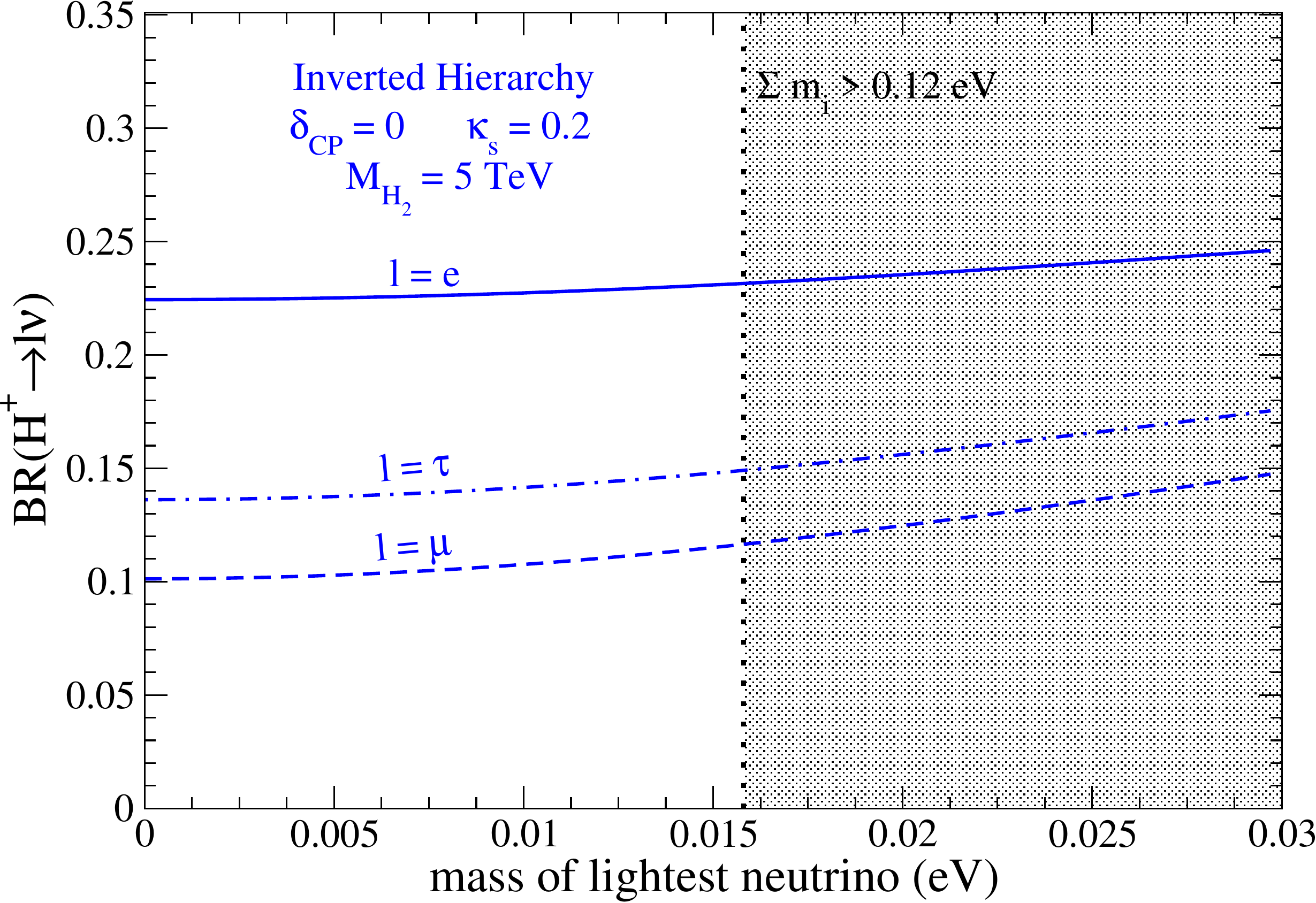}\label{fig:BR_inv}}\\
\subfigure[]{\includegraphics[width=0.45\textwidth,clip]{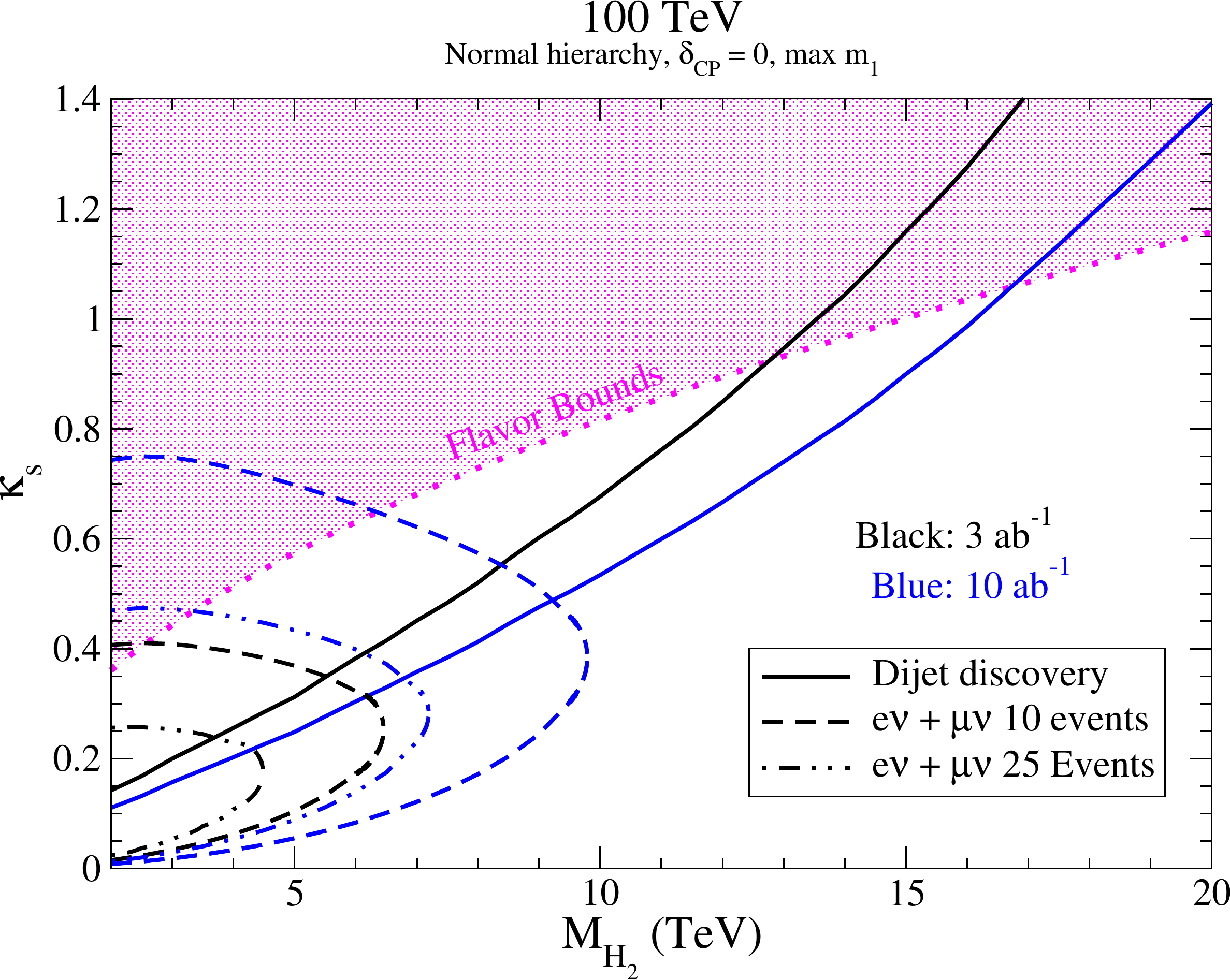}\label{fig:discovery_a}}
\subfigure[]{\includegraphics[width=0.45\textwidth,clip]{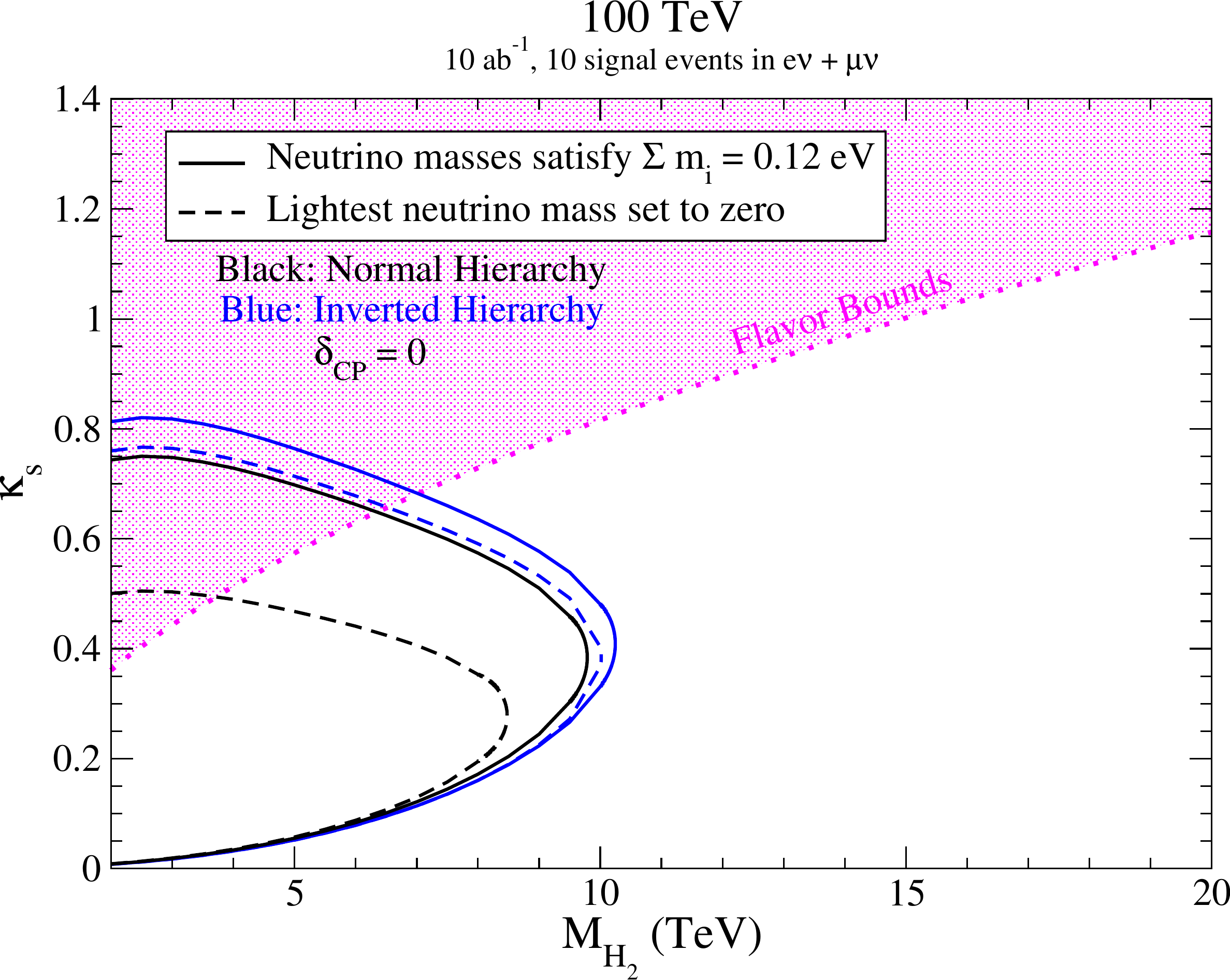}\label{fig:discovery_b}}
\caption{\label{fig:StrangeMass} (a,b) Branching ratios of charged Higgs into leptonic final states in the (a) normal and (b) inverted hierarchies with $\delta_{CP}=0,\,\kappa_s=0.2,\,$ and $M_{H_2}=5$~TeV.  For the inverted hierachy we show when the cosmological bound on the sum of neutrino masses $\sum m_i<0.12$~eV~\cite{Ivanov:2019hqk} is violated in the gray shaded region.  (c,d) Discovery reaches at a 100 TeV $pp$ machine with flavor bounds in the shaded magenta region.  (c)~Discovery reach in (solid) dijet searches and $e\nu$ and $\mu\nu$ finals states with 10 and 25 events within the dashed and dot-dot-dashed lines, respectively.  (d) Discovery reaches in $e\nu$ and $\mu\nu$ final states with 10 events and 10 ab$^{-1}$ of data for the (black) normal and (blue) inverted hierarchies when the lightest neutrino mass is (dashed) zero or (solid) saturates the cosmological bound $\sum m_i<0.12$~eV~\cite{Ivanov:2019hqk}.  The components of the heavy Higgs doublet are assumed to be degenerate in mass.}  
\end{center}
\end{figure*}

In this model we work in a two Higgs doublet model, but with the same flavor framework as Eqs.~(\ref{eq:uptypeSFV},\ref{eq:lamnu}).  That is, we use Model B as detailed above.  When the Higgs bosons are integrated out, the operators in Eq.~(\ref{eq:Yukawa}) will give rise to a dimension-6 operator of the form $\bar{q}q\bar{\nu}\nu$, where $q$ are quarks and $\nu$ are neutrinos (with left-right pairing).  When QCD condenses, the $\bar{q}q$ develops a vacuum expectation value, which will give neutrino masses.  Although the QCD vacuum expectation value is small, we have shown~\cite{Davoudiasl:2021nfv} that this is a viable scenario to generate small neutrino masses with order one coupling to the heavy Higgs and a Higgs mass $\ord{10~{\rm TeV}}$.

The operator $\bar{q}q\bar{\nu}\nu$ can mediate many light meson decays, depending on the quark flavor.  Adapting the results of Ref.~\cite{Davoudiasl:2005ai}, the partial decay widths of mesons into an electron plus neutrino via this operator are~\cite{Davoudiasl:2021nfv}
\beq
\Gamma(P^+ \to e^+ \, \nu_R) = \frac{\sum_i |\zeta_{ei}|^2}{64 \pi\, M_D^4} f_P^2 \mu_P^2 m_P\,,
\label{Gamma}
\eeq
where for the mesons $P=\pi,K,D_S$,
\beq
\mu_P=\frac{m_\pi^2}{2 \bar m}\quad;\quad \frac{m_K^2}{m_s + \bar m}\quad;\quad \frac{m_{D_s}^2}{m_{c} + m_s}\,,
\label{muP}
\eeq
$M_D$ is the mass of the heavy Higgs, $\zeta_{ei}$ is the Wilson coefficient of $\bar{q}q\bar{\nu}\nu$, $\bar{m}=(m_u+m_d)/2$, $m_{u,d,s,c}$ are quark masses, and $m_{\pi,K,D_S}$ are meson masses.

There are stringent bounds on how large the meson branching ratios can be~\cite{Zyla:2020zbs}\footnote{We have checked~\cite{Davoudiasl:2021nfv} that similar results can be obtained via $R$-ratios.}:
\begin{eqnarray}
&{\rm Br}(\pi^+\rightarrow e^+\,\nu)&=\,(1.23\pm0.004)\times 10^{-4}\\
&{\rm Br}(K^+\rightarrow e^+\,\nu)&=\,(1.582\pm 0.007)\times 10^{-5}\\
&{\rm Br}(D_s^+\rightarrow e^+\,\nu)&<\, 8.3\times 10^{-5}\quad \text{(90\% C.L.).}\label{BRs}
\end{eqnarray}
The bounds on pions and kaons are very constraining, and it is not possible to have order one couplings and generate the requisite neutrino masses with heavy resonance masses below $\sim 60$~TeV.  These masses are much too large to be able to generate the neutrino masses.  However, the $D_s$ bounds are much less constraining.  Indeed, for order one couplings, the bounds can be satisfied for $M_D\gtrsim 3$~TeV, which could be observable at future colliders.

From this discussion, we focus on a scenario in which the the heavy Higgs couples only to strange quarks and leptons.  The operator under consideration is then
\beq
O_D = \zeta\frac{[\bar Q\,s]\epsilon[\bar L \, \nu_R]}{M_D^2} + {\small \rm H.C.}\,,
\label{OD}
\eeq
where $Q$ denotes the second generation quark doublet $(c,s)_L$.  
When the strange quark condensation 
$\scond \approx -(300~\text{MeV})^3$ \cite{Davies:2018hmw} is considered, we find that viable neutrino masses $m_\nu\sim0.1$~eV can be obtained with $M_D\sim 16$~TeV and $\zeta\sim 1$.  This mass is consistent with the rare meson decay constraints.

To generate neutrino masses, we can solve for the Yukawa couplings of the neutrinos in terms of the strange quark Yukawa, heavy Higgs mass, strange quark condensate, and neutrino masses.  The neutrino mass eigenstate couplings are then
\begin{eqnarray}
\kappa_{\nu,i}=\frac{M_{H_2}^2}{\kappa_s\,\langle \bar{s}s\rangle}m_i.\label{eq:kapnu}
\end{eqnarray}
As this makes clear, the couplings of the flavor eigenstates to the heavy Higgses will depend strongly on the neutrino mass scale and mass ordering.

In Figs.~\ref{fig:StrangeMass}(a) and (b) we show the branching ratios of the heavy charged Higgs into lepton final states as a function of the lightest neutrino mass for the (a) normal and (b) inverted hierarchies for a representative parameter point.  As is clear, there can be substantial branching ratios into leptonic final states for reasonable parameters.  The hierarchy of branching ratios depends strongly on the neutrino mass hierarchy: for the normal hierarchy the branching ratio into muons dominates while for the inverted the branching ratios into electrons dominates.  Finally, for the normal hierarchy, the branching ratios depend fairly strongly on the neutrino mass scale.  {\it Hence, if this model is realized in Nature, measuring the decay rates of  heavy charged Higgs scalars into leptons can provide insight into the neutrino mass hierarchy and scale.}

In Fig.~\ref{fig:StrangeMass}(c) we show, for the normal hierarchy with maximum neutrino masses, the discovery reach for a 100 TeV $pp$ collider in the (solid line) dijet and (within the dashed, dot-dot-dashed lines) lepton plus missing energy channels.  The black lines are assuming 3 ab$^{-1}$ of data and the blue lines 10 ab$^{-1}$.   We overlay the relevant flavor bounds in the shaded magenta region.  Since the only non-zero quark coupling are to $s_R$,  the relevant flavor bounds come from $D-\bar{D}$ mixing in Fig.~\ref{fig:BaryoFlavor}.  As is clear, the dijet and leptonic searches cover complementary parameter regions with the dijets more sensitive at higher Higgs masses and the leptonic searches more relevant at lower Higgs masses.

The different 5$\sigma$ discovery  in the leptonic channels for (blue) inverted and (black) normal hierarchies at a 100 TeV $pp$ collider are shown in Fig.~\ref{fig:StrangeMass}.  These reaches are for  10 events with 10 ab$^{-1}$ of data and for the lightest neutrino mass set to (within the solid lines) its maximum and (within the dashed lines) zero.  The inverted hierachy is more easily discovered, due to the larger branching fractions into muons and electrons as shown in Figs.~\ref{fig:StrangeMass}(a,b). 

\section{Conclusions}

The presence of multiple Higgs doublets in extensions of the SM is a reasonable possibility.  In particular, if electroweak symmetry is mainly broken by the vev of the Higgs field corresponding to the observed boson at $\sim 125$~GeV, one may ask how the extra scalars can play a role in addressing the shortcomings of the minimal SM.  In the above, we summarized two proposals that address this question.  In either case, the new heavy scalars have $\ord{1}$ couplings to light quarks.  To avoid severe constraints from FCNCs, the ``up-type Spontaneous Flavor Violation" framework was used in both cases and adapted to the structure of the models.      

In one scenario, two heavy extra Higgs doublets were added to the SM in order to provide a possible mechanism for baryogenesis, through CP violating decays of the new heavy scalars and production of a lepton asymmetry.  If the SM neutrinos are Dirac particles this model can provide an alternative to conventional leptogenesis,  which relies on heavy right-handed neutrinos and generically leads to light Majorana neutrinos.  A future 100 TeV $pp$ collider can have reach up to $\sim 20-30$~TeV, through resonant production and decay of the scalars into di-jets.  Depending on the choice of couplings to down-type quarks, current flavor data, in particular from $D-\bar D$ mixing, yield important constraints on the parameters of the model.  If the scalars are sufficiently degenerate in mass, allowing coherently enhanced signals, the LHC may be able to probe this model up to masses of $\sim 5-7$~TeV.

In a second proposal, we considered the unconventional possibility that Dirac neutrino masses may originate from the quark condensate of the SM QCD.  This model requires one extra Higgs doublet that has significant coupling to a light quark.  Precision measurements of charged meson decays require that this coupling mainly involve the second generation quarks, with the strange quark condensate sourcing the requisite electroweak symmetry breaking.  Due to the constrained nature of the required effective operator that mediates Dirac neutrino mass generation, the new scalar masses in this model are not expected to be heavier than $\sim 10-15$~TeV.  The new scalars can be produced on resonance through their quark couplings and be discovered in di-jet final states, at a 100 TeV $pp$ collider with $\lsim 10$~ab$^{-1}$ of integrated luminosity.  Remarkably, measurements of the charged Higgs branching ratios into leptons can furnish a probe of the mass hierarchy of the SM neutrinos, in this model.  Again, existing $D-\bar D$ mixing data provide  significant constraints on the parameters of the scenario.


\begin{acknowledgments}
{\bf Acknowledgments:} The work of H.D. and M.S. is supported by the United States Department of Energy under Grant Contract DE-SC0012704.  I.M.L. is supported in part by the United States Department of Energy grant number DE-SC0017988.  The data to reproduce the plots are available upon request.
\end{acknowledgments}

\bibliography{main}

\end{document}